\documentclass[reprint,amsmath,amssymb]{revtex4-1}
\usepackage[utf8x]{inputenc}
\usepackage{graphicx}
\usepackage{dcolumn}
\usepackage{bm}
\usepackage{color}

\usepackage{soul} 
\usepackage[normalem]{ulem} 

\begin{document}

\preprint{APS/123-QED}

\title{Calculating the polarization in bi-partite lattice models:
  application to an extended Su-Schrieffer-Heeger model}

\author{Bal\'azs Het\'enyi$^{1,2}$, Yetkin Pulcu$^1$, and Serkan
Do\v{g}an$^3$}
\affiliation{$^1$ Department of Physics, Bilkent
  University, TR-06800 Bilkent, Ankara, Turkey \\ $^2$MTA-BME Quantum
  Dynamics and Correlations Research Group, Department of Physics,
  Budapest University of Technology and Economics, H-1111 Budapest,
  Hungary \\ $^3$ Department of Mathematics, Bilkent University,
  TR-06800 Bilkent, Ankara, Turkey }

\date{\today}

\begin{abstract}
We address the question of different representation of Bloch states
for lattices with a basis, with a focus on topological systems.  The
representations differ in the relative phase of the Wannier functions
corresponding to the diffferent basis members.  We show that the phase
can be chosen in such a way that the Wannier functions for the
different sites in the basis both become eigenstates of the position
operator in a particular band.  A key step in showing this is the
extension of the Brillouin zone.  When the distance between sites
within a unit cell is a rational number, $p/q$, the Brillouin extends
by a factor of $q$.  For irrational numbers, the Brillouin zone
extends to infinity.  In the case of rational distance, $p/q$, the
Berry phase ``lives'' on a cyclic curve in the parameter space of the
Hamiltonian, on the Brillouin zone extended by a factor of $q$.  For
irrational distances the most stable way to calculate the polarization
is to approximate the distance as a rational sequence, and use the
formulas derived here for rational numbers.  The use of different
bases are related to unitary transformations of the Hamiltonian, as
such, the phase diagrams of topological systems are not altered, but
each phase can acquire different topological characteristics when the
basis is changed.  In the example we use, an extended
Su-Schrieffer-Heeger model, the use of the diagonal basis leads to
toroidal knots in the Hamiltonian space, whose winding numbers give
the polarization.
\end{abstract}

\pacs{}

\maketitle

\section{Introduction}

Recently, several papers investigated~\cite{Bena09,Watanabe18} the
different representation of the wave function and physical observables
when the underlying model is a lattice with a basis.  This fundamental
question is important {\it per se}, moreover, such models have been
the subject of intense research in the last decades, exemplified by
graphene, the kagome, or Lieb lattices, and others.  The canonical
models for topological insulators~\cite{Bernevig13,Kane13,Asboth15}
are all lattices with bases: the Haldane,~\cite{Haldane88}
Kane-Mele,~\cite{Kane05a,Kane05b} Su-Schrieffer-Heeger~\cite{Su79}
models are some of the most fundamental ones.

Bena and Montambaux~\cite{Bena09} have studied the two commonly used
bases of the graphene lattice.  The difference between the two is that
in one (basis $I$), the different position of the sites within a unit
cell are not explicit (they are taken to be the same, the position of
the unit cell itself), whereas in the other (basis $II$), the
positions of the different sites within a unit cell are explicit.
What is meant by this is that in $I$, the Wannier functions
corresponding to the different basis members have the same phase,
while in $II$ their relative phase depends on the distance between
them.  Bena and Montambaux~\cite{Bena09} show that in the case of the
graphene lattice, although the Hamiltonian and operators take
different forms in the two bases, important physical quantities, such
as the density or the density of states, as well as the low-energy
theory, are unchanged.  However, care must be taken in using the
correct construction of observables, once a representation of the wave
functions is chosen.

Watanabe and Oshikawa~\cite{Watanabe18} have taken up this question
more recently, and they show that, while the precise form of the
current operator does depend on the choice of basis, the quantized
charge pumped during an adiabatic cycle~\cite{Thouless82} is
unchanged.  Note that the current operator is intimately related to
the phase of the wave function, in particular the relative phase wave
functions centered on different sites may exhibit.  Also note that the
charge pumped is a topological invariant, which remains the same in
different bases.

In this paper we reconsider the question of basis dependence in
lattice models in the context of modern polarization theory with a
focus on topological systems.  We compare the two bases used in
Ref. \cite{Bena09}, and show that one corresponds to a linear
combination of Wannier functions which are eigenfunctions of the
position operator.  When the distance between the basis sites is a
rational number $p/q$, the Brillouin zone has to be extended by a
factor of $q$ to derive the above result.  If the distance is
irrational, then the Brillouin zone has to be extended to the whole
real line.

We then turn to the question of topological phase transitions in the
two bases.  As an example system we consider an extended
Su-Schrieffer-Heeger~\cite{Su79} (SSH) model.  When the $k$-space
Hamiltonian is written considering distance dependence explicitly the
three-dimensional curve $d_x,d_y,d_z$ (see Eq. \ref{eqn:H}) is not a
closed curve.  When the distance between basis sites is a rational
number ($p/q$) extending the Brillouin zone by a factor of $q$ closes
the curve, and an expression for the polarization Berry
phase~\cite{Berry84,Zak89,King-Smith93,Resta98,Resta00} can be
derived.  In the irrational case a closed $d_x,d_y,d_z$-curve results
only if the Brillouin zone is extended to infinity.  For this reason,
calculation of the irrational case is not straightforward.  We show
that a stable calculation results if the irrational distance is
approximated as a rational sequence.

Implementing the dependence of the Hamiltonian on the distance between
basis sites proceeds through a unitary transformation.  As such, the
Hamiltonian can acquire auxiliary topological features.  In the model
we study, the $d_x,d_y,d_z$-curve traces out a toroidal knot, and the
ratio of the winding numbers of the knot give the polarization.
However, the phase diagram itself is not altered, since these
topological characteristics are merely different representations of
the phases of the original extended SSH model.

In the next section we discuss the two different bases used in
lattices with a basis.  In section \ref{sec:model} the extended SSH
model is described.  In section \ref{sec:pol} the change of basis is
discussed in the context of the extended SSH model.  In section
\ref{sec:rslts} our results are presented, and in section
\ref{sec:cnclsn} we conclude our work.

\section{Polarization and choice of representation}
\label{sec:polbas}

It is well-known~\cite{Blount62,Kivelson82} that Wannier functions in
systems without a basis are eigenstates of the position operator
within a given band.  Here we show that in a lattice with a basis, one
of the representations of the wave-function studied by Bena and
Montambaux~\cite{Bena09} leads to position eigenstates.

We consider a one-dimensional system whose potential obeys
$V(x)=V(x+1)$ (in other words we take the length of a unit cell as
unity).  The unit cell consists of a two-site basis.  We chose the
distance between the two internal sites to be $p/q$, a rational
number, and we extend the Brillouin zone by a factor of $q$.  This
step is justified below.

In this case, the two commonly used bases~\cite{Bena09} can be written
as
\begin{eqnarray}
  \nonumber
  u_k^I(x) =& \sum\limits_{m=-\infty}^{\infty} e^{i k (m-x)}\left[w_a(x-m) + w_b\left(x-m-\frac{p}{q}\right)\right] \\   \nonumber
  u_k^{II}(x) =& \sum\limits_{m=-\infty}^{\infty} \left[e^{i k (m-x)} w_a(x-m) + e^{i k (m+p/q-x)}\times\right. \\
   &\left.  \times w_b\left(x-m-\frac{p}{q}\right)\right].
\end{eqnarray}
The functions $u_k^I(x)$ and $u_k^{II}(x)$ are periodic Bloch
functions.  $u_k^I(x)$ is the more standard approach, and it
appears~\cite{Ashcroft76} in textbooks.  $w_a(x),w_b(x)$ are two
different localized Wannier functions.  In the extended SSH model we
study below, they differ only in having their centers displaced with
respect to each other.

Following Zak~\cite{Zak89} we derive the geometric phase which arises
upon integrating the connection across the Brillouin zone.  We define
a geometric phase of the Zak type, in the extended Brillouin zone as:
\begin{equation}
  \gamma = \frac{i}{2\pi q} \int_{-q\pi}^{q\pi} dk \langle u_k
  |\partial_k| u_k \rangle.
\end{equation}
Using $u_k^I$ the Zak phase becomes
\begin{equation}
  \gamma_I = \int_{-\infty}^\infty d x x \left|w_a(x) + w_b\left(x-\frac{p}{q}\right)\right|^2,
\end{equation}
whereas with $u_k^{II}$ 
\begin{equation}
  \gamma_{II} = \int_{-\infty}^\infty dx x \left(|w_a(x)|^2 +
  |w_b(x)|^2 \right).
\end{equation}
In deriving $\gamma_{II}$, we made use of the fact that
\begin{equation}
  \label{eqn:deltan}
  \frac{1}{2\pi q} \int_{-\pi q}^{\pi q} d k \exp\left(i k (n-n')/q
  \right) = \delta_{n n'}.
\end{equation}
Comparing $\gamma_I$ and $\gamma_{II}$ we see that the second Bloch
basis function, $u_k^{II}$ is a diagonal basis for the position
operator in a periodic system, since the result is a simple sum of
Wyckoff positions for the two different Wannier functions.  It is due
to Eq. (\ref{eqn:deltan}) that the cross terms in $\gamma_{II}$
disappear, but only of the Brillouin zone is extended by a factor of
$q$.

It is instructive to consider the effect of inversion symmetry on
either $\gamma$.  In the original work of Zak~\cite{Zak89}, where the
sites are equivalent, two types of inversion symmetries were
considered, site-inversion and bond-inversion.  The former gives a Zak
phase of zero, while the latter gives $\pi$, which corresponds to a
polarization of $q a /2$ ($a$ is the length of the unit cell).  In our
case, there are also two types of inversion.  If the sites of the
model are a distance $p/q$ apart, one can consider inverting around
the midpoint of the $p/q$ bond, or the $1-p/q$ bond.  In the former
case it holds that $w_c(-x+p/q)=\pm w_c(x)$ for both $c=a,b$, which
leads to $\gamma = \frac{p}{2q}$ (for both $I$ and $II$).  In deriving
this one can follow the stame steps as Zak\cite{Zak89}.  Inverting
around the other bond center leads to $w_c(-x-1+p/q)=\pm w_c(x)$, from
which it follows that $\gamma = \frac{p-q}{2q}$ (again for both $I$
and $II$).

In the case of irrational distance, the Brillouin zone has to be
extended to the range $(-\infty,\infty)$.  When this is done,
Eq. (\ref{eqn:deltan}) eliminates the cross terms and leads to
$\gamma_{II}$.

\section{Model}
\label{sec:model}

The full model we consider is an extension of the well-known
Su-Schrieffer-Heeger model~\cite{Su79}.  It is a bi-partite lattice
model ($A$ and $B$ sublattices), whose Hamiltonian has the form,
\begin{equation}
  \label{eqn:H_c}
  \hat{H} = - \sum_{i=1}^L [ Jc_i^\dagger d_i +  J'd_i^\dagger c_{i+1} + i K c_i^\dagger c_{i+1} - i K d_i^\dagger d_{i+1} + \mbox{H.c.}],
\end{equation}
where $L$ denotes the number of unit cells, $c_i$($c_i^\dagger$)
denote the annihiliation(creation) operators on the $A$ site of the
$i$th unit cell, and $d_i$($d_i^\dagger$) denote the same for the $B$
site of the $i$th unit cell.  We take the length of a unit cell to be
unity.  An extended SSH model of a different kind, with real second
nearest neighbor hoppings, was invesigated by Li et al.~\cite{Li14}.
  \\

When $K=0$ the SSH model is recovered.  The third and fourth terms in
Eq. (\ref{eqn:H_c}) correspond to second nearest neighbor hoppings
(hoppings within one sublattice).  A Peierls phase of $\pi/2$ and
$-\pi/2$ is applied along these hoppings, the former for sublattice
$A$, the latter for sublattice $B$.  These second nearest neighbor
hoppings play a similar role to those in the Haldane
model~\cite{Haldane88}.  There and in Eq. (\ref{eqn:H_c}) the Peierls
phases of second nearest neighbor hoppings point in opposite
directions on the two sublattices.  The consequence of this term is
that for gapped phases and finite $K$ there is persistent current in
the system, but the currents on the different sublattices cancel, and
insulation results.  If $K$ changes sign, the direction of the
persistent currents on the different sublattices also change sign.  In
this sense, the model can be viewed as an example of Kohn's
tenet~\cite{Kohn64}.  This tenet states that insulation in a quantum
system is not a function of localization of individual charge
carriers, but ultimately results from many-body localization, or the
localization of the center of mass of all the charge carriers.
Individually, charge carriers can be quite mobile, as long as the
center of mass is localized.\\

\begin{figure}[ht]
 \centering
 \includegraphics[width=8cm,keepaspectratio=true]{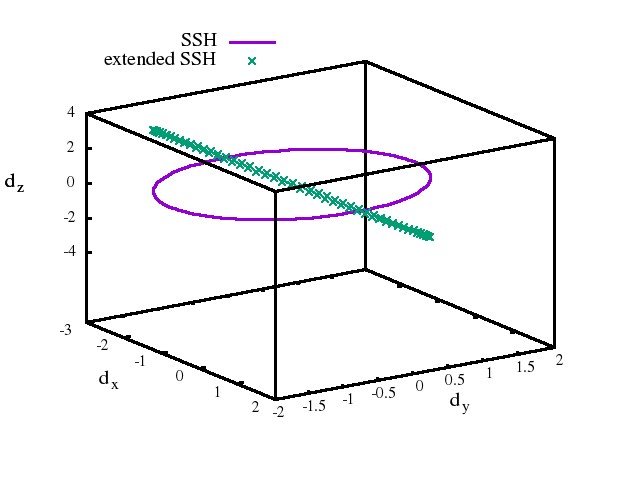}
 \includegraphics[width=8cm,keepaspectratio=true]{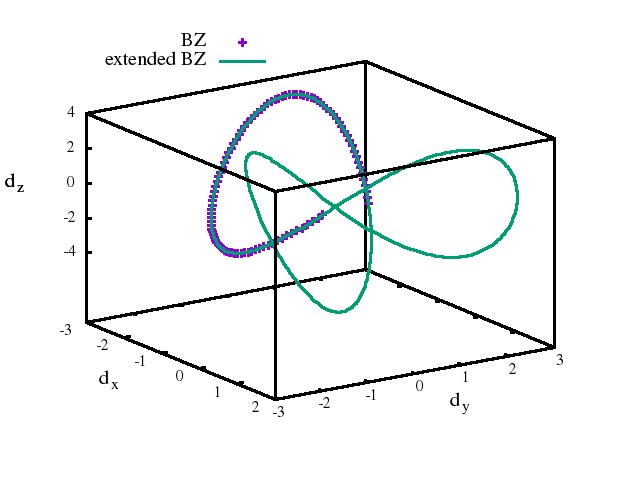}
 \caption{Curves traced out by the parameters of the Hamiltonian
   $d_x,d_y,d_z$ as $k$ traverses Brillouin zone for the usual SSH and
   the extended SSH model.  Upper panel: SSH model and extended SSH
   model in the usual basis.  Lower panel: extended SSH model in the
   distance dependent basis comparing the first Brilllouin zone to the
   extended Brillouin zone.}
 \label{fig:SSH}
\end{figure}

According to the symmetry classification~\cite{Altland97,Schnyder08}
of topological insulators, the SSH model, which exhibits time
reversal, particle-hole, and chiral symmetries, falls into the BDI
class, which in one dimension is a $\mathbb{Z}$-topological system.
After the addition of the second nearest neighbor hopping term in
Eq. (\ref{eqn:H}) only the particle-hole symmetry remains, and the
model falls in the D symmetry class, which is a $\mathbb{Z}_2$ system.
The gap closure points of the SSH remain (they occur at $k=0,\pi$),
and we show below that topological edge states still arise.  \\

Fourier transforming Eq. (\ref{eqn:H_c}) results in
\begin{equation}
  \label{eqn:H}
  \hat{H} = \sum_k H_k = \sum_k[ d_x(k) \sigma_x + d_y(k) \sigma_y + d_z(k) \sigma_z],
\end{equation}
where $\sigma_x, \sigma_y, \sigma_z$ denote the Pauli matrices, and
\begin{eqnarray}
  d_x(k) &=& - J - J' \cos(k),\\ \nonumber
  d_y(k) &=& - J' \sin(k),  \\ \nonumber
  d_z(k) &=&  2 K \sin(k).
\end{eqnarray}
Written this way, we see that gap closure occurs at $k=\pi$ if $J=J'$.
The upper panel Fig. \ref{fig:SSH} shows the curves traced out by
$d_x,d_y,d_z$ as $k$ traverses the Brillouin zone for the usual SSH
model ($J=1.25,J'=1.75$) with $K=0$ and an example of the extended
one, with $K=2$.  The former forms a circle in the $xy$-plane, the
latter is a circle with a tilted axis (axis along the $yz$-plane).
Both circles cross the $x$-axis itself in two places at the same
points.  This means, that by varying the ratio of $J/J'$ it is
possible to close the gap in both models, and encounter topological
phase transitions.

We can also develop a continuum approximation, the usual
way~\cite{Jackiw76},
\begin{equation}
  \label{eqn:H_cont}
  H = i t \partial_x \sigma_x + i K \partial_x \sigma_y + 2 \delta t
  \sigma_z,
\end{equation}
where we used the parametrization:
\begin{eqnarray}
  \label{eqn:prm}
  t &=& \frac{J + J'}{2} \\
  \delta t &=& \frac{J'-J}{2} \nonumber
\end{eqnarray}
We can solve for zero energy eigenstates of the form
\begin{equation}
  \begin{pmatrix}
    \Psi_A(x) \\ \Psi_B(x)
  \end{pmatrix}.
\end{equation}
The zero energy solution is
\begin{equation}
  \Psi_{A/B}''(x) = \frac{4 \delta t^2}{ t^2 + K^2} \Psi_{A/B}(x).
\end{equation}
The solutions are exponentials.  If the system is extended, these
solutions diverge, hence they are unnormalizable.  If the system has
open boundary conditions, exponential solutions can be normalized,
since they can start at one of the boundaries and decay towards the
bulk.  Therefore zero-energy states are possile.  The sign of $\delta
t$ determines the direction of decay, in other words, the side of the
finite system on which the edge state is located.  The signs of $t$
and $K$ are irrelevant.

\begin{figure}[ht]
 \centering
 \includegraphics[width=8cm,keepaspectratio=true]{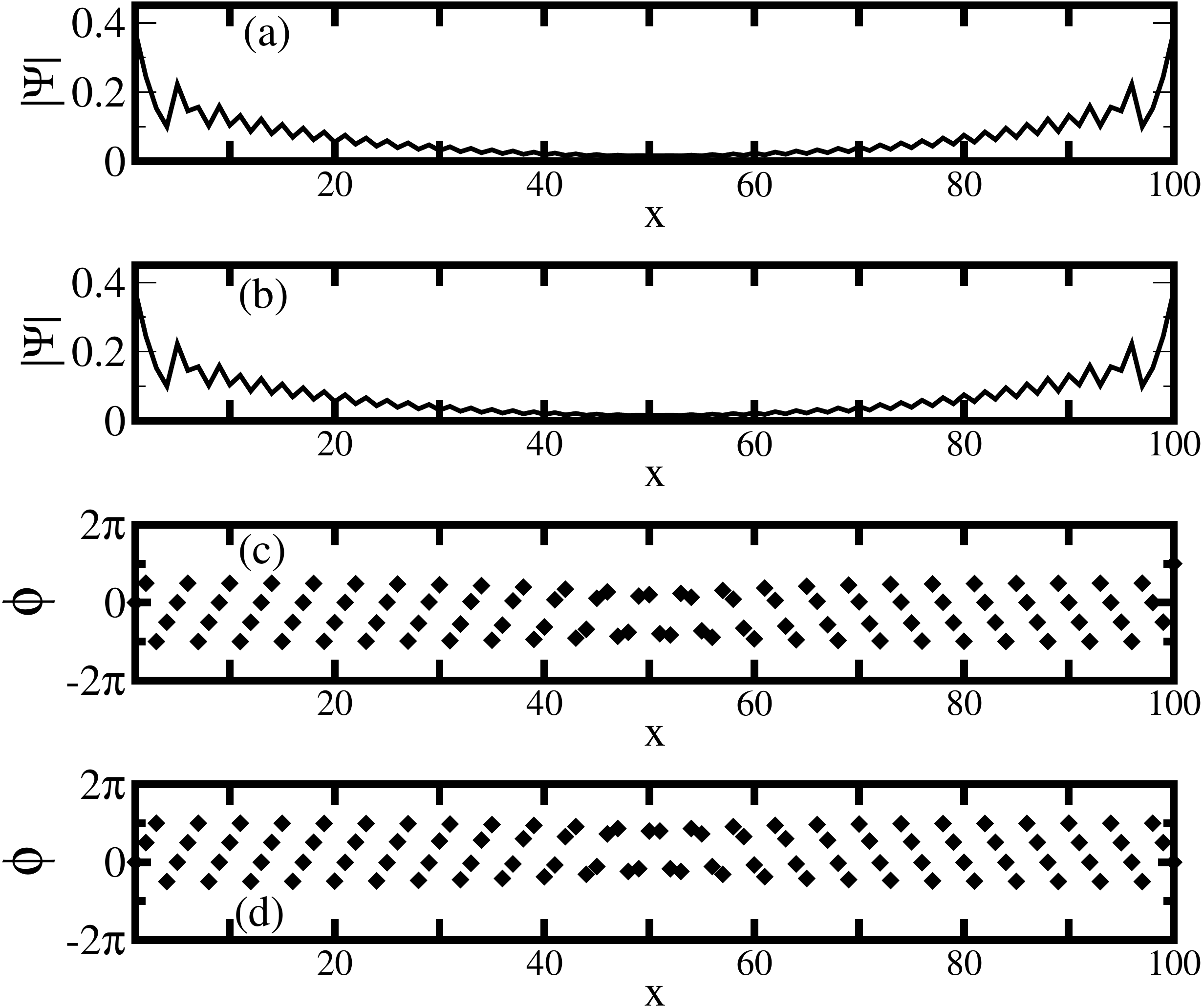}
 \caption{Calculation for a system with open boundary conditions with
   $L=100$ (meaning also that there are $100$ states).  The parameters
   of the Hamiltonian are $J=1.25, J'=1.75, K=2.0$.  The minimum
   distance between sublattices is $p/q=1/3$.  The figure show: (a)
   the zero energy state number $50$, (b) the zero energy state number
   $51$, (c) the phase of the wave function as a function of lattice
   site for state number $50$, and (d) the phase of the wave function
   as a function of lattice site for state number $51$.}
 \label{fig:edge_state}
\end{figure}

It is interesting to compare the above results to those of Jackiw and
Rebbi~\cite{Jackiw76}.  There a first order differential equation is
solved for a function multiplying a $\sigma_z$ eigenstate, while in
our case each component of the Pauli spinor satisfies the same second
order differential equation.  Our calculations below on the lattice
model (see Fig. \ref{fig:edge_state}) bear out the predictions of our
continuum theory.

\section{Implementation of distance dependence}

\label{sec:pol}

We take the shortest distance between the two sublattices to be $p/q$,
where $p$ and $q$ are co-prime integers.  The Hamiltonian in $k$-space
can be shown to be:
\begin{eqnarray}
  \label{eqn:H_pq}
  d_x(k) &=& - J \cos\left[ k \left( \frac{p}{q}\right) \right]
  - J' \cos\left[ k \left(1-\frac{p}{q} \right) \right] \\ \nonumber
  d_y(k) &=&  J \sin\left[ k \left( \frac{p}{q}\right) \right]
  - J' \sin\left[ k \left(1-\frac{p}{q} \right) \right] \\ \nonumber
  d_z(k) &=&  2 K \sin(k).
\end{eqnarray}
This Hamiltonian can also be derived from the original one
(Eq. (\ref{eqn:H})) by
\begin{equation}
  \label{eqn:rot}
H_k \rightarrow  \exp\left(i k \frac{\sigma_z}{2} \frac{p}{q} \right) H_k \exp\left(-i k \frac{\sigma_z}{2} \frac{p}{q} \right),
\end{equation}
a rotation by angle $k p/q$ around the $z$ axis.  We also note that the
vector function $[d_x,d_y,d_z]$ has a period of $ 2 \pi q$,
rather than the usual $2\pi$.

Due to the fact that the distance dependence coincides with a
rotation, a qualitative difference between the case of rational and
irrational distances arises.  As is known~\cite{Asboth15,Kane13}, the
curve traced out by the SSH model on the $d_x-d_y$ plane is an ellipse
whose center, in general, lies on the $d_x$ axis.  Implementing the
distance dependence will generate a different curve.  A closed curve
can be obtained if the distance between sublattice sites is taken to
be a rational number ($p/q$ with $p$ and $q$ coprimes), and the
Brillouin zone is extended to $ 2 \pi q$.  Observables related to the
Berry (Zak) phase (polarization, topological invariants) depend on the
curve being closed, since these physical quantities are integrals over
a connection over a closed curve (usually the Brillouin zone).
\begin{figure}[ht]
 \centering
 \includegraphics[width=5cm,keepaspectratio=true]{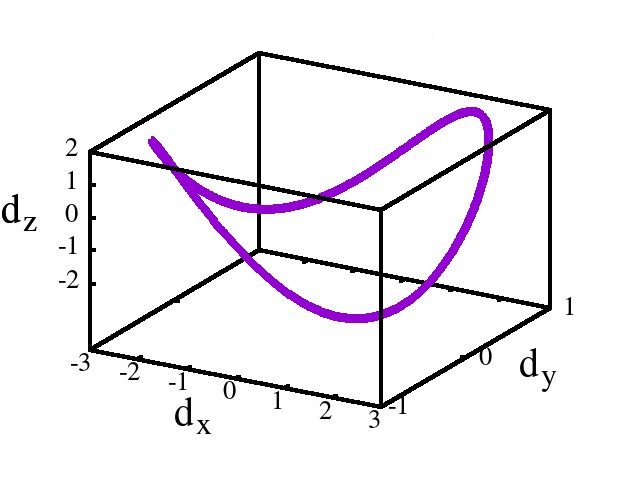}
 \includegraphics[width=5cm,keepaspectratio=true]{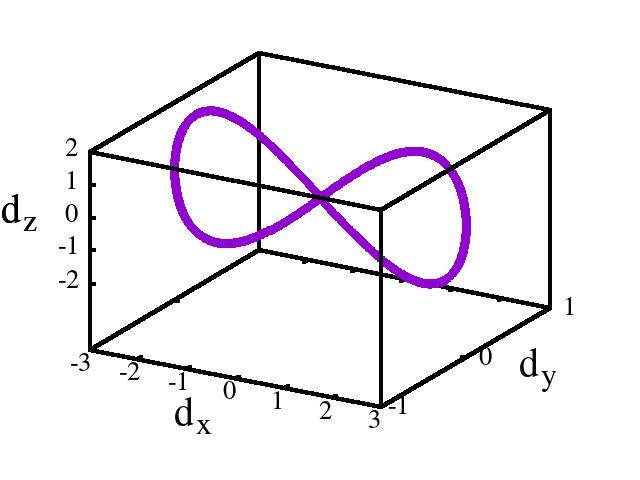}
 \includegraphics[width=5cm,keepaspectratio=true]{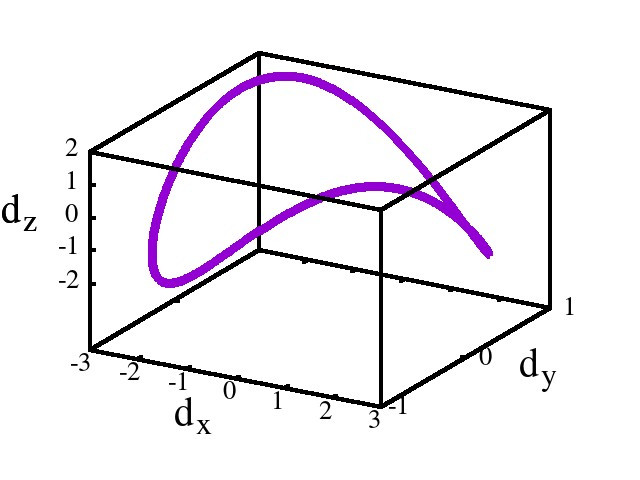}
 \caption{Curves traced out by the parameters of the Hamiltonian
   $d_x,d_y,d_z$ as $k$ traverses the extended Brillouin zone
   for the $\frac{p}{q}=\frac{1}{2}$ case.  The Hamiltonian parameters
   are $K=1$ in all cases, and $J=1,J'=2$ (upper panel),
   $J=1.5,J'=1.5$ (middle panel), $J=2,J'=1$ (lower panel).  The
   curves of the upper and lower panels appear equivalent, but the
   orientation of the curves changes as the gap closure point (the
   figure-8 of the middle panel) is crossed.}
 \label{fig:p1q2}
\end{figure}

Let us rewrite the $d$-curve of Eq. (\ref{eqn:H_pq}) with a scaled
wave vector, $k = q \kappa$, as
\begin{eqnarray}
  \label{eqn:H_pq_tknot}
  d_x(\kappa) &=& - \cos(\kappa p) \left[ J + J' \cos(\kappa q) \right] \\ \nonumber
              & & - \sin(\kappa p) J \sin(\kappa q), \\  \nonumber
  d_y(\kappa) &=& + \sin(\kappa p) \left[ J + J' \cos(\kappa q) \right] \\  \nonumber
              & & - \cos(\kappa p) J' \sin(\kappa q), \\  \nonumber
  d_z(\kappa) &=& + 2K \sin (\kappa q).  \nonumber
\end{eqnarray}
This curve exists on the surface of a torus, albeit, not a torus of
the usual parametrization, and the curve itself is a toroidal knot
(see examples in Figs. \ref{fig:p1q2}, \ref{fig:pXq3},
\ref{fig:pXq5}).

The fundamental group~\cite{Nakahara03} of the torus is $\pi_1(T^2) =
\pi_1(S^1) \oplus \pi_1(S^1) = \mathbb{Z} \oplus \mathbb{Z}$, meaning
that homotopically equivalent classes of loops on the torus can be
characterized by two integers.  In Eq. (\ref{eqn:H_pq_tknot}) the two
integers are $p,q$, and in this parametrization they are toroidal
winding numbers.  A torus knot is obtained by a trajectory which loops
around the hole of the torus an integer number of times ($w_1=p$),
while it makes an integer number of rotations around the ``body'' of
the torus itself ($w_2=q$).  As argued in the previous section, the
ratio of winding numbers is proportial to the polarization.  A curve
parametrized as Eq. (\ref{eqn:H_pq_tknot}) traces out a toroidal knot
for a given integer pair $p,q$, even if $p$ and $q$ are not co-primes.
If they are co-primes then there is only one curve, if they are not,
then the same curve is traced more than once, as many times as the
common divisor of $p$ and $q$.  In the original parametrization,
Eq. (\ref{eqn:H_pq}), the rescaling leads to the same curves for a
given $p,q$.

Alternatively, one can define $r = q - p$ in Eq. (\ref{eqn:H_pq}), and
arrive at a different curve, similar in form to
Eq. (\ref{eqn:H_pq_tknot}).  In this case the winding numbers will be
$w_1=r$ and $w_2=q$.  Thus, we can write the two different
fractionally quantized polarization results of the previous section
as:
  \begin{equation}
    \label{eqn:Pw1w2}
  \mathfrak{P} = -\frac{e}{2}\frac{w_1}{w_2}.
\end{equation}

The lower panel of Fig. \ref{fig:SSH} shows two curves for this
extended model, one using a normal Brillouin zone ($-\pi,\pi$), the
other an extended one ($-q\pi,q\pi$) (model parameters are
$J=1.25,J'=1.75,K=2$ and $p/q=1/3$ in both cases).  The former curve
is open, the latter is a closed curve, a toroidal knot.

In our calculations below the Brillouin zone is stretched by a factor
of $q$, and we take every $q$th point in $k$ space in this enlarged
Brillouin zone.  The starting point for the polarization expression is
the scalar product
\begin{equation}
  Z_q = \langle \Psi |\exp\left(i \frac{2 \pi \hat{X}}{L}q \right)|
  \Psi \rangle.
\end{equation}
Following the steps of Resta~\cite{Resta98}, in a band insulator we
have
\begin{equation}
  \label{eqn:Zqband}
  Z_q = \prod_{s=0}^{L-1} \mbox{det} S(k_{qs},k_{q(s+1)}).
\end{equation}
where
\begin{equation}
  S_{m,m'}(k_{qs},k_{q(s+1)}) = \int_0^L dx \hspace{.1cm} u^*_{k_{qs},m}(x)
  u_{k_{q(s+1)},m'}(x),
\end{equation}
where $u_{k_{qs},m}(x)$ denote the periodic Bloch functions of band
$m$, and $k_{qs} = 2 \pi q s/L$.  In our numerics we calculate the
polarization via
\begin{equation}
  \label{eqn:P}
  \mathfrak{P} = -\frac{e}{2 \pi} \mbox{Im} \ln \prod_{s=0}^{L-1}
           [\mbox{det} S(k_{qs},k_{q(s+1)})]^{\frac{1}{q}}.
\end{equation}
This definition is important from the point of view of implementation,
because the $\frac{1}{q}$ power keeps the terms in the Berry phase
between $0$ and $2\pi$, and provides the correct fractional
polarization.

Again the question remains, whether irrational distances can be
handled.  The difficulty in this case is that the Brillouin zone has
to be extended to infinity.  Two methods suggest themselves.  One, use
the irrational distance itself in place of $p/q$, and compare larger
and larger Brillouin zones.  In this case, the curve in the
$d_x,d_y,d_z$ curve will be open.  Another way would be to approximate
the irrational distance as the limit of a sequence of rational
numbers, and investigate the limit of Eq. (\ref{eqn:P}) along that
sequence.  Below a comparison is presented for the case when the
distance is the inverse of the golden ratio. \\

\section{Results}
\label{sec:rslts}

The curves traced out by the parameters of the Hamiltonian
($d_x,d_y,d_z$) are shown in Fig. \ref{fig:p1q2} for three
cases: $J=1,J'=2$; $J=1.5,J'=1.5$; $J=2,J'=1$ ($K=1$ in all three
cases) for $p=1,q=2$.  The topological transition occurs at
$J=1.5,J'=1.5$, where the curve traces out a figure$-8$, with a
crossing point at the origin.  The curves on the different sides of
this phase diagram (the uppermost and lowermost panels) can be
transformed into each other via a rotation by $\pi/2$.  Since the
orientation of the curves (the sense of rotation around the
$d_z$-axis) does change when crossing the phase boundary the winding
number goes from one to minus one.  This corresponds to a polarization
reversal from $\mathfrak{P} = e/4$ to $\mathfrak{P} = -e/4$, as
expected.

\begin{figure}[ht]
 \centering
 \includegraphics[width=8cm,keepaspectratio=true]{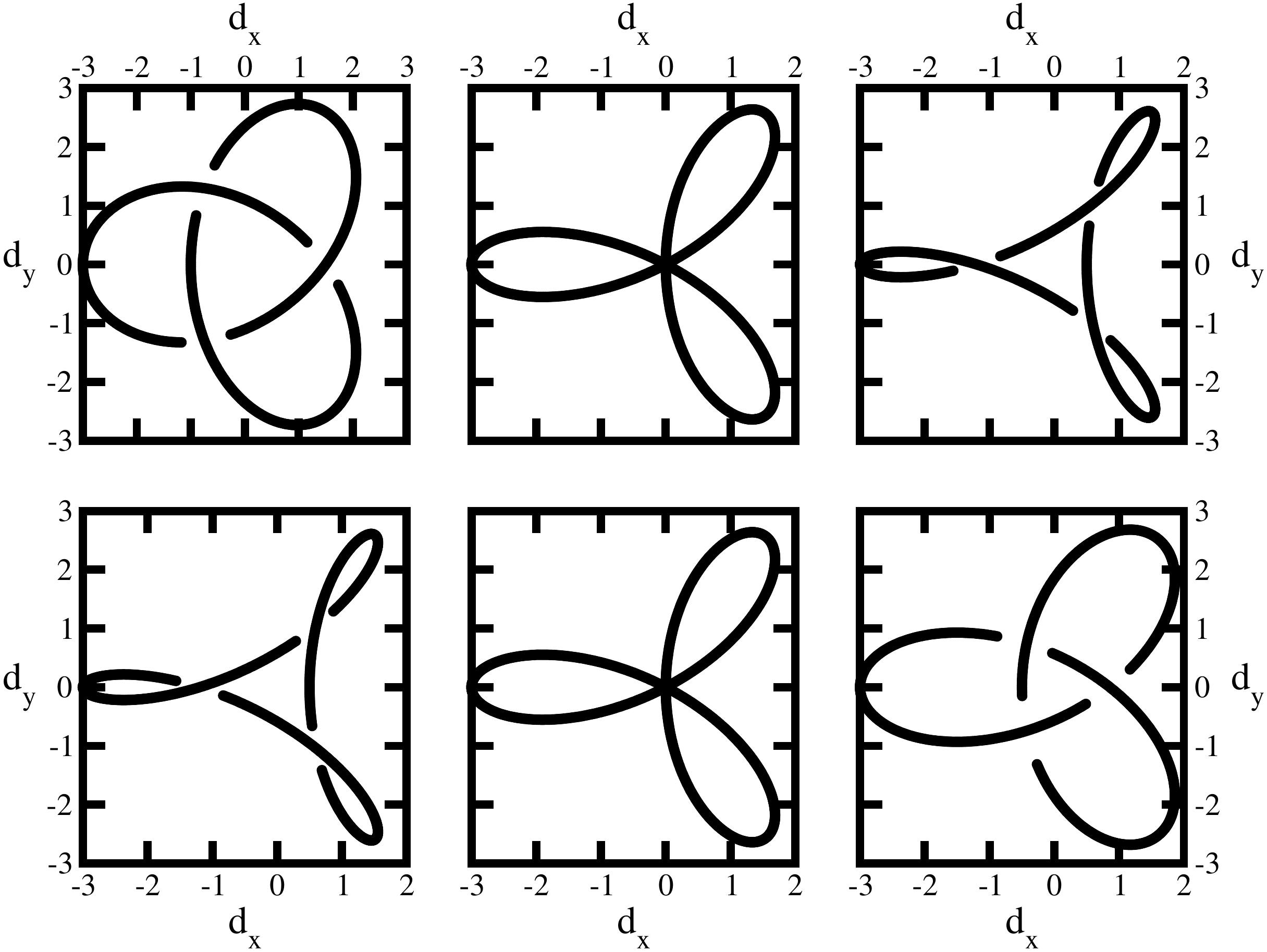}
 \caption{Curves traced out by the parameters of the Hamiltonian
   $d_x,d_y,d_z$ as $k$ traverses the extended Brillouin zone.  The
   upper(lower) set of plots show the case
   $\frac{p}{q}=\frac{1}{3}$($\frac{p}{q}=\frac{2}{3}$), where the
   parameters of the Hamiltonian are, left: $J'=1.75,J=1.25$,
   center:$J=1.5,J'=1.5$, and right:$J'=1.25,J=1.75$ ($K=1$ in all
   cases).}
 \label{fig:pXq3}
\end{figure}

Fig. \ref{fig:edge_state} presents a study of how edge states arise in
a system with open boundary conditions.  The parameters of the
Hamiltonian are $J=1.25, J'=1.75, K=2.0$ with $p=1$ and $q=3$.  The
upper two panels ((a) and (b)) show the two zero energy wave functions
which are localized at the edges.  The system is fairly small $L=100$,
and the paramater $K$ is fairly large, giving rise to edge state
wavefunctions with a sizable finite component even halfway through the
lattice, however, as the size is increased and/or if $K$ is decreased
the bulk compoment of the state decreases.  These results are
consistent with our previous derivation (Eq. (\ref{eqn:H_cont})).  For
a system with $J'<J$ edge-states are not found.  The bottom panels
((c) and (d)) show the phase of the zero energy wave functions.  The
phases of both edge states oscillate when going towards the center
from the edges.  The left and right sides mirror each other, and a
sudden change occurs at and within a few sites near the center of the
system.

\begin{figure}[ht]
 \centering
 \includegraphics[width=8cm,keepaspectratio=true]{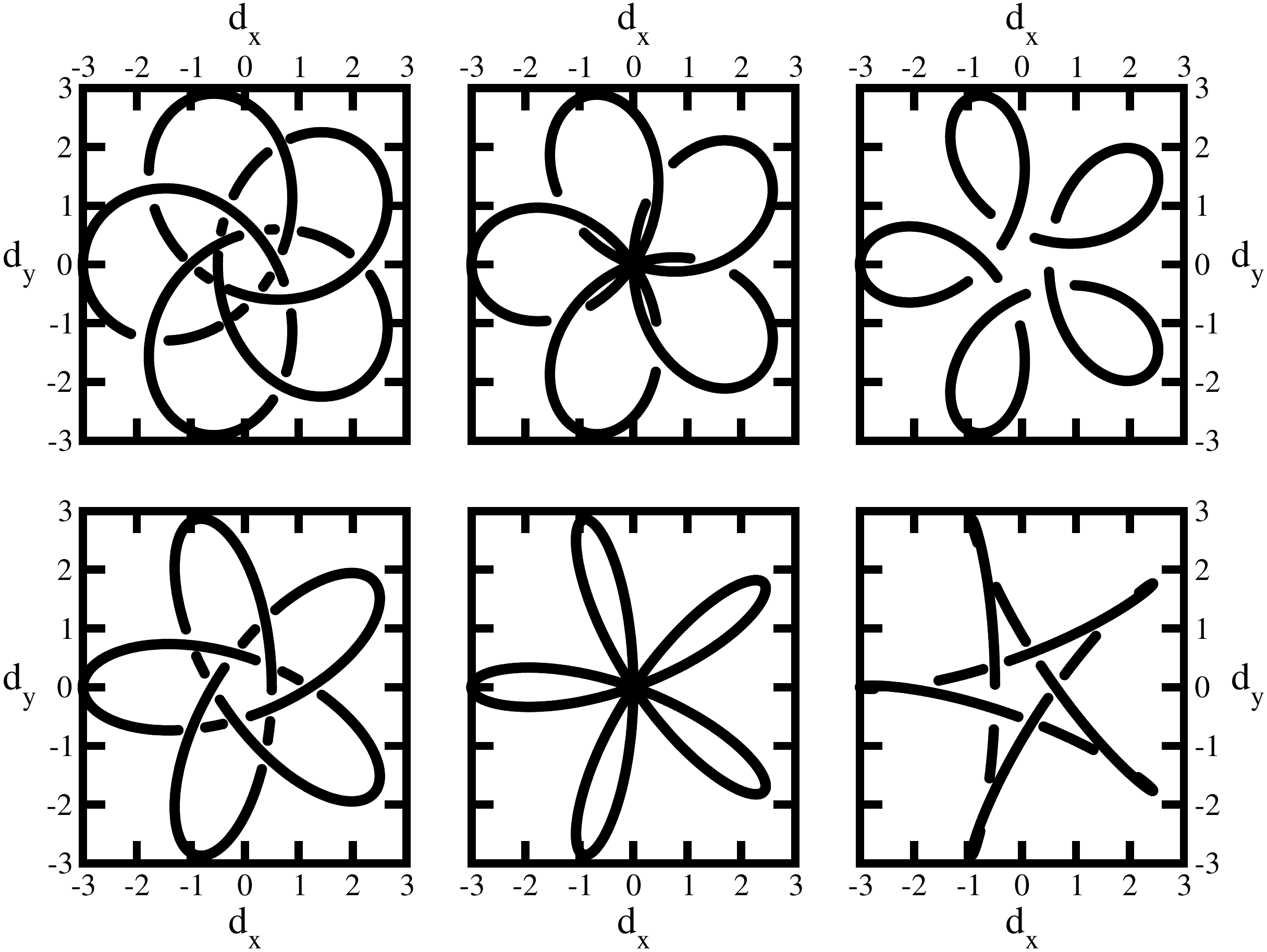}
 \caption{Curves traced out by the parameters of the Hamiltonian
   $d_x,d_y,d_z$ as $k$ traverses the extended Brillouin zone.  The
   upper(lower) set of plots show the case
   $\frac{p}{q}=\frac{1}{5}$($\frac{p}{q}=\frac{2}{5}$).  The
   parameters of the Hamiltonian are, left: $J'=1.75,J=1.25$,
   center:$J=1.5,J'=1.5$, and right:$J'=1.25,J=1.75$ ($K=1$ in all
   cases).}
 \label{fig:pXq5}
\end{figure}

In Figs. \ref{fig:pXq3} and \ref{fig:pXq5} the $d_x,d_y,d_z$ curves
are shown for various values of $p$ and $q$.  The Hamiltonian
parameters are as follows: left $J'=1.75,J=1.25$, center
$J'=1.5,J=1.5$, and right $J'=1.25,J=1.75$.  $K=1.0$ for all cases.
The plots are knot diagrams~\cite{Adams01}, two dimensional knot
representations, arrived at by projecting the curve onto the $d_x-d_y$
plane, but indicating which strand is below the other at crossings.
Other than the plots in the center in both figures, neither of which
are knots, the left and right plots show torus knots.  For example, in
the top left figure of Fig. \ref{fig:pXq5}, the curve winds around
$w_1=4$ times while it makes $w_2=5$ revolutions around the body of
the torus.

The center plots in both Figs. \ref{fig:pXq3} and \ref{fig:pXq5} show
curves for the topological phase transition.  These curves do not form
knots, they cross at the origin, which is the gap closure point.  It
is obvious that these center plots ``connect'' the two topologically
distinct phases on either side of the phase transition.  At the phase
transition the torus surface on which the curves ``live'' is a horn
torus.  On either side of the horn torus, the curves reconnect, in a
manner reminiscent of ``partner switching'' of edge states in the
Kane-Mele model~\cite{Kane05a,Kane05b}.

\begin{figure}[ht]
 \centering
 \includegraphics[width=8cm,keepaspectratio=true]{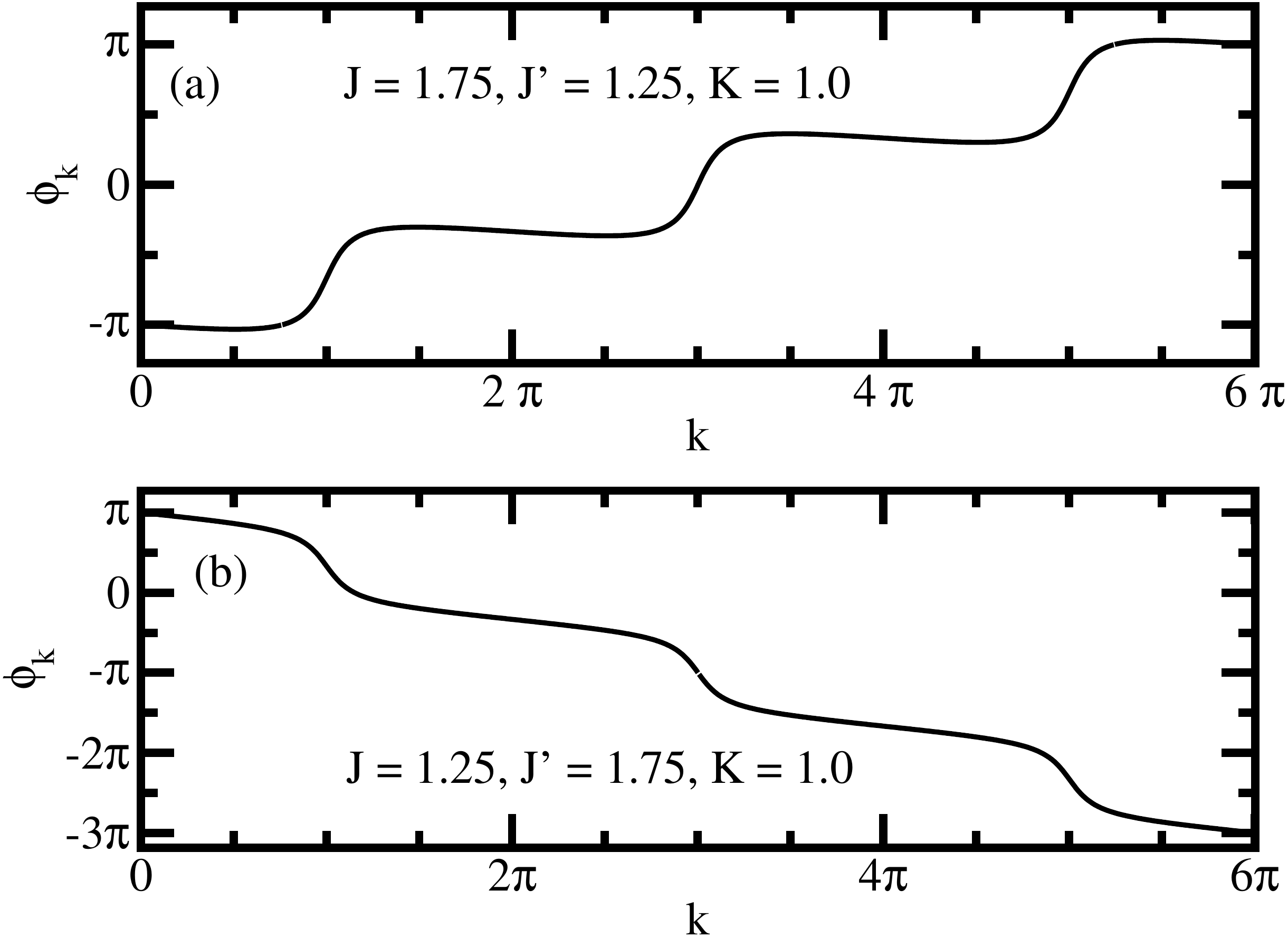}
 \caption{The trajectory of $\phi_k$ a parameter which defines the
   wave-function (Eq. (\ref{eqn:wf})) across the extended Brillouin
   zone for a trivial (a) and a topological (b) case.  As $k$ sweeps
   across the extended Brillouin zone $\phi_k$ increases by $2 \pi$ in
   the trivial, and $-4\pi$ in the topological system.}
 \label{fig:winding}
\end{figure}

The upper plots in Fig. \ref{fig:pXq3} are for $p=1$, the lower ones
are $p=2$.  The upper left plot shows a trefoil knot with $w_1=2$,
$w_2=-3$.  The upper right plot shows an un-knot with $w_1=1$, $w_2=3$.
Even though there are crossings on this curve, they can be eliminated
by type I Reidemeister moves~\cite{Adams01}, one of three
modifications on two-dimensional representations of knots which leave
knot invariants unaltered.  The knot invariants (in this case the two
winding numbers) are not altered by these modifications (if all
three twists are eliminated).

The topological phase transition connects the two states.  In the
insulating phases Eq. (\ref{eqn:Pw1w2}) is recovered.  In the lower
set the $J'>J$ curve exhibits the unknot ($w_1=1, w_2=-3$), and the
$J>J'$ the trefoil knot ($w_1=2, w_2=3$), again consistent with
Eq. (\ref{eqn:Pw1w2}).

In Fig. \ref{fig:pXq5} the $d_x,d_y,d_z$ curves are shown for various
$q=5$ cases.  The upper plots are $p=1$, the lower ones are $p=2$.
The phase transition in the upper set is a topological one connecting
a $w_1=4$ phase ($J'>J$) with a $w_1=1$ phase ($J>J'$), while the
bottom set shows a transition between a $w_1=3$ phase ($J'>J$) and a
$w_1=2$ phase ($J>J'$), both accompanied by sign changes in $w_2$
whose absolute value is five.  The cases $p=3$ and $p=4$ (not shown)
give winding numbers of $w_1=2$, $w_1=3$ and $w_1=1$, $w_1=4$,
respectively, again with changes in sign in $w_2$ across the quantum
phase transition.  In all cases, our polarization formula,
Eq. (\ref{eqn:Pw1w2}), is recovered.

To summarize the information of Figs. \ref{fig:pXq3} and
\ref{fig:pXq5}, it is useful to connect the results to what is known
from the ``usual'' representation~\cite{Su79}, in which the Wannier
functions corresponding to different basis sites are taken to have the
same phase (basis $I$).  There, there are two distinct phases, one
with winding number zero (trivial phase), one with winding number one
(topological phase), separated by gap closure.  Performing the unitary
transformation in Eq. (\ref{eqn:rot}) leads to a different
representation of the same topological phases, and it is these
different representations which are shown in the figures.  In a given
system with fixed $p/q$, the insulating phases are knots on ring or
spindle tori.  A ring torus can be deformed into a spindle torus via a
horn torus.  The horn torus occurs if $J=J'$, that is when the gap
closes, and where the phase transition separating different
topological phases occurs (center panels in Figs. \ref{fig:pXq3} and
\ref{fig:pXq5}).  Note that in the $p/q=1/3$ case, the
topological(trivial) phase corresponds to the spindle(ring) torus,
whereas the reverse is true for $p/q=2/3$.

When $K=0$ (the usual SSH model), the above curves remain on the
$d_x,d_y$ plane, hence one cannot speak of a torus.  In the ``usual''
representation~\cite{Asboth15,Kane13} of this model the $J'>J$ is is
the topological state with winding number of one, while the $J'<J$ is
the trivial site with winding number zero, depending on whether the
$d_x,d_y$ curve encircles the origin.  Implementing distance
dependence, turns each of those phases into phases with distinct
quantized polarizations.  Gradually changing the sign of $K$ changes
the sign of both winding numbers for a given toroidal knot (which goes
into its chiral counterpart), leading to no change in the polarization
(since the polarization is the ratio of winding numbers).  This is
consistent with the result derived in section \ref{sec:model}: the
sign of $K$ does not determine the presence of edge states (nor
whether the state is topological or trival), only the sign of the
$J'-J$ does.

\begin{figure}[ht]
 \centering
 \includegraphics[width=8cm,keepaspectratio=true]{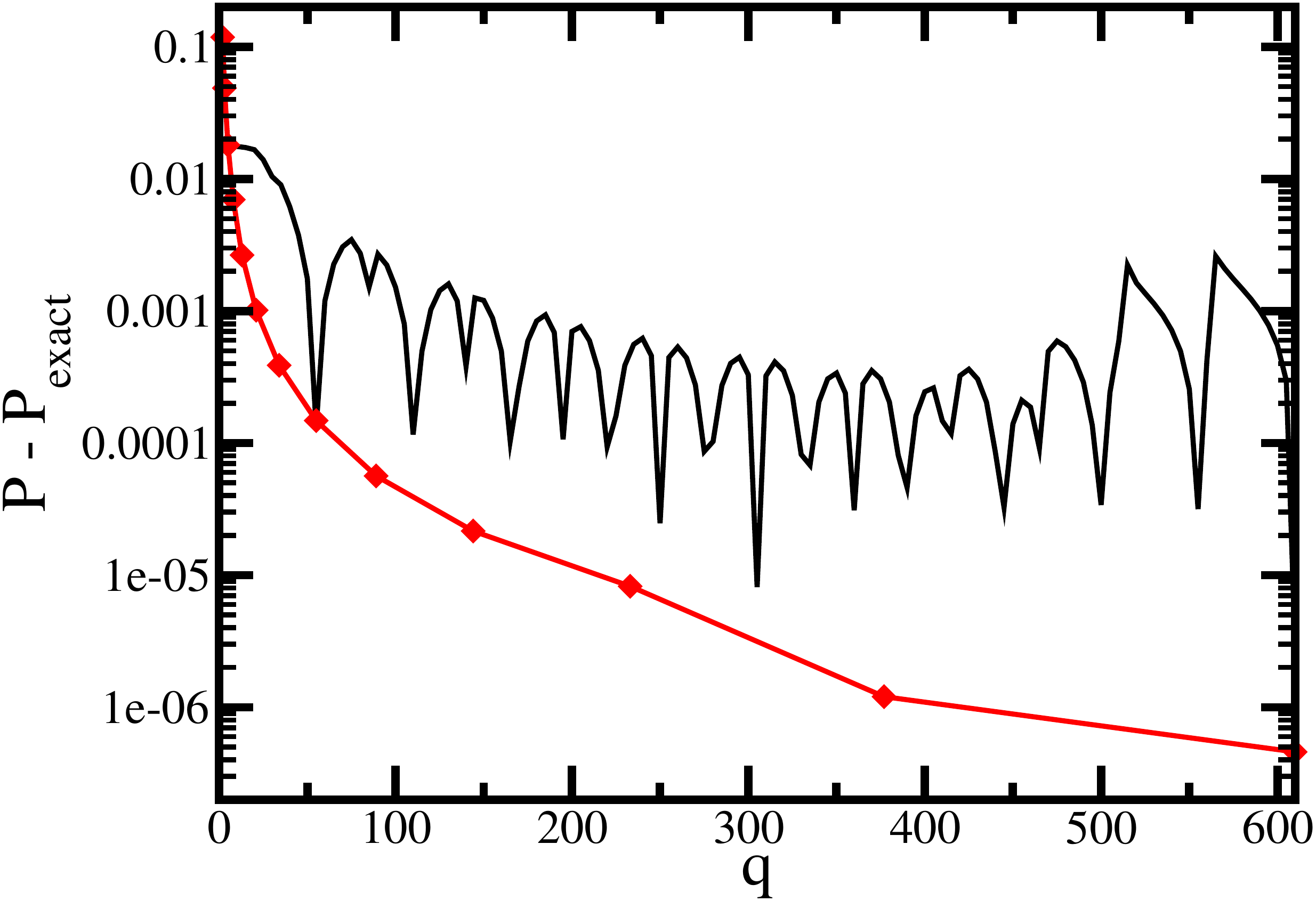}
 \caption{Semilogarithmic plot of the absolute value of the difference
   between the exact value of the polarization $P_{exact}$ and the
   polarization calculated via two different methods.  The solid like
   shows the use of an irrational distance (inverse of the golden
   ratio), the diamonds show a calculation based on a rational
   approximation to the inverse golden ratio.}
 \label{fig:DeltaP}
\end{figure}

Another way of seeing how the fractional polarization arises is to
look at the phase variable of the wave function.  The wave functions
that are the solution of the Hamiltonian in Eq. (\ref{eqn:H}) with the
$d$-components given by Eq. (\ref{eqn:H_pq}) can be written in the
form
\begin{equation}
  \label{eqn:wf}
  \begin{pmatrix}
    \sin \left( \frac{\theta_k}{2} \right) \\ \cos \left( \frac{\theta_k}{2} \right) \exp (i \phi_k)
  \end{pmatrix},
\end{equation}
where $\theta_k$ and $\phi_k$ depend on $d_x,d_y,d_z$
(Eq. (\ref{eqn:H_pq})).  The phase $\phi_k$ is shown in
Fig. \ref{fig:winding}, for a system with $p=1$ and $q=3$.  In the
trivial phase, the phase $\phi_k$ changes by $2\pi$ as $k$ covers the
extended Brillouin zone ($2 \pi q = 6 \pi$).  Accordingly, the
polarization is $\mathfrak{P} = - \frac{e}{2}\frac{1}{3}$.  Meanwhile,
in the topological phase, the variable $\phi_k$ changes by $-4\pi$
along the Brillouin zone, and the polarization is $\mathfrak{P} =
\frac{e}{2}\frac{2}{3}$.  

In Fig. \ref{fig:DeltaP} we present calculations for an irrational
distance, namely the inverse of the golden ratio.  In one calculation,
we used the irrational number itself to derive a Bloch Hamiltonian (in
place of $p/q$ in Eq. (\ref{eqn:H_pq})), and Brillouin zones of size
$-\pi q, \pi q$ were used.  In another calculation ratios of the
Fibonacci sequence were used to approximate the golden ratio, and
Eq. (\ref{eqn:P}) was used.  The results in Fig. \ref{fig:DeltaP} show
that, while the polarization based on using an irrational distance is
reasonably close to the exact result, using a rational approximation
is a much more stable procedure, which uniformly approaches the exact
result.
  
\begin{figure}[ht]
 \centering
 \includegraphics[width=8cm,keepaspectratio=true]{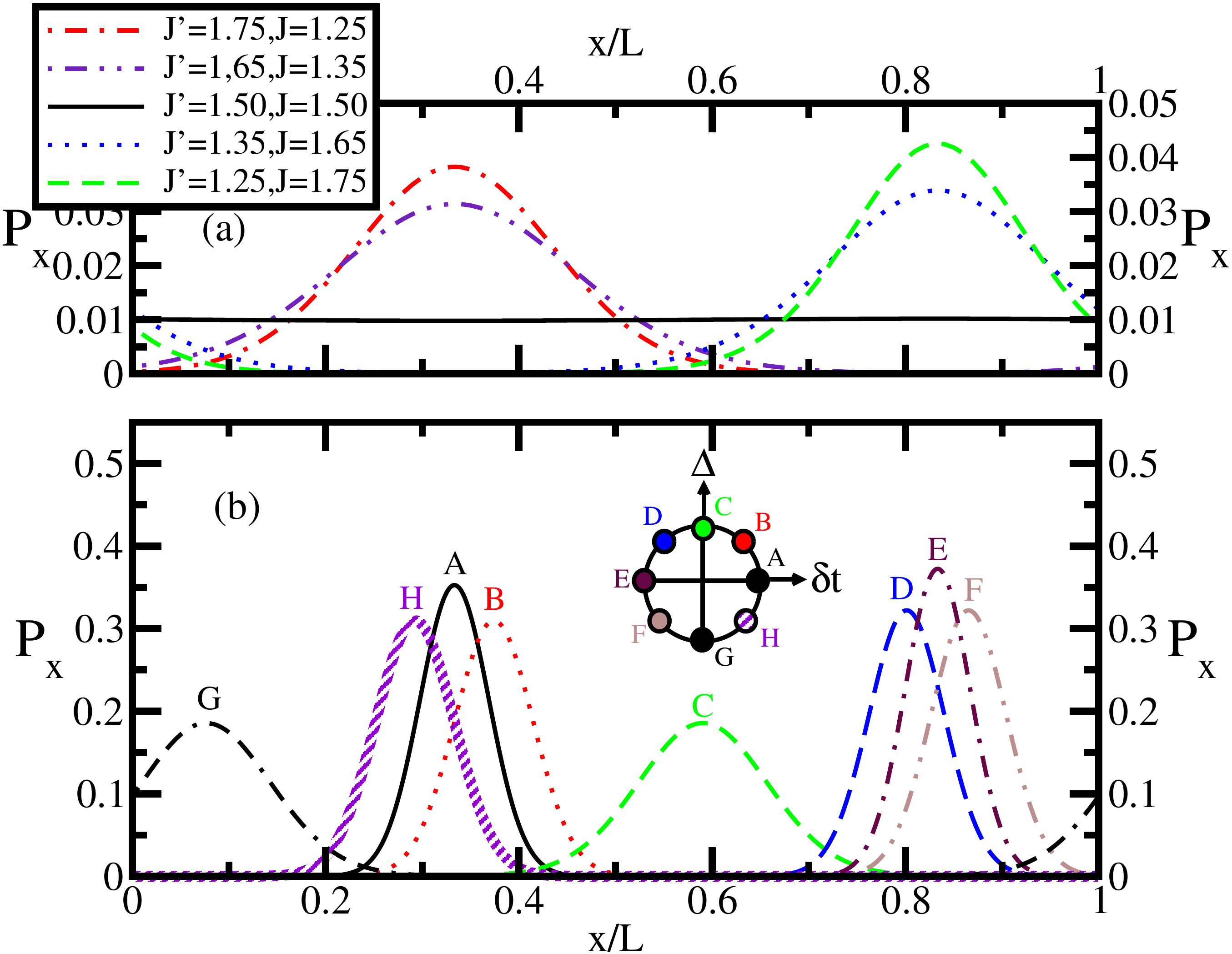}
 \caption{(a) Fourier transform of the polarization amplitude, or
   polarization distribution shown for different cases of the system
   with $p/q=1/3$; two sets in each ordered state and at the gap
   closure point.  The curves are normalized to one over the lattice
   of $L=100$.  Legend is valid only for the upper panel.  (b)
   Reconstructed polarization distributions along an adiabatic charge
   pumping process.  For further explanation see the text.}
 \label{fig:phastrans}
\end{figure}

If an adiabatic charge pumping experiment~\cite{Atala13,Nakajima16}
undergoing a {\it full} cycle is considered the amount of charge
pumped would be one unit of charge per cell.  This coincides with the
results of Watanabe and Oshikawa~\cite{Watanabe18} who showed that the
charge pumped is independent of how the Hamiltonian is written.
Fig. \ref{fig:phastrans}a shows the Fourier transform of $Z_q$,
defined as
\begin{equation}
  P_x = \sum_q \exp(i 2 \pi x q /L) Z_q; \hspace{.2cm} x = 1,...,L;
\end{equation}
which corresponds to the underlying polarization
distribution~\cite{Souza00,Resta99,Yahyavi17,Hetenyi19}.  The case
$p/q=1/3$ is considered.  Shown are two different parameter sets for
each ordered state, as well as the phase transition, where $J=J'$.
The distributions of the ordered states show maxima corresponding to
the values of the polarization calculated above.  The tuning of the
parameters of the Hamiltonian within one phase change the shape of the
distribution (variance and other cumulants), but not the position of
the maximum.  At the transition the distribution becomes nearly flat,
the system does not exhibit a well-defined polarization due to gap
closure.  For a finite system the curve representing $J=J'$ is not
entirely flat, because the $k-$space sampling does not precisely hit
the points where the gaps occur.  However, as the thermodynamic limit
is taken, the curve progressively flattens out.

In the cases treated heretofore the polarization can be easily
determined by the ratio of the winding numbers, hence the Berry phase
formula is not necessary.  If other terms are added to the
Hamiltonian, for example, an on-site potential, then the Berry phase
based formula in Eq. (\ref{eqn:P}) is needed.  In the lower panel of
Fig. \ref{fig:phastrans}b an adiabatic charge pumping
process~\cite{Nakajima16} is shown.  The Hamiltonian is reparametrized
according to Eq. (\ref{eqn:prm}) and an alternating on-site potential
of strength $\Delta$ is also added.  In the calculations of
Fig. \ref{fig:phastrans}b $t=K=1$.  The plots shown are along total
polarization distribution curves along the circle in the $\delta
t-\Delta$ plane.  The equation of the circle is
$0.2\cos(\alpha),0.2\sin(\alpha)$, and the curves labeled
$A,B,C,D,E,F,G,H$ correspond to values of $\alpha = 0, \pi/4, \pi/2,
3\pi/4, \pi, 5\pi/4, 3\pi/2, 7\pi/4$ respectively.  As $\alpha$
increases, the maximum of the polarization distribution function
shifts towards the right within the unit cell.  For the curves $G$ and
$H$ the maxima of the curves is to the left of the $A$-curve.  This
can be interpreted as a charge which entered from a neighboring unit
cell (to the left of the one shown), as is expected to occur for
adiabatic charge pumping.  Note that at points $A$ and $E$ the maxima
correspond to the ones shown in Fig. \ref{fig:phastrans}a, and the
maximum of the charge distribution appears to travel smoothly in
between.  The part of the adiabatic charge pump process between
$\alpha=0$ and $\alpha=\pi$ occurs between systems which exhibit
fractionally quantized polarization.  If the $I$ basis was used, the
maximu would be located at the edge of and halfway through the unit
cell.

\section{Conclusion}
\label{sec:cnclsn}

Lattices with a basis can be solved by various representations of the
wave function.  The representations differ in the relative phases of
the Wannier functions associated with different sites within the unit
cell.  The different representations are related by unitary
transformations.  The transformation can be used to diagonalize the
position operator, if t he Brillouin zone is extended.  When the
nearest distance between basis sites is a rational number $p/q$, the
Brillouin zone has to be extended by a factor of $q$.  When this
distance is an irrational number, the Brillouin zone extends to the
entire real line.  This latter state of affairs raises the question of
how to calculate the polarization, to which we suggested the use of a
rational approximation to the distance between intra-unit cell sites.

Applying such a transformation to the Hamiltonian can give rise to
auxiliary topological features.  In our example calculations, based on
an extension of the SSH model, the $d_x,d_y,d_z$-curve forms a
toroidal knot, and the ratio of the two winding numbers of the torus
is proportional to the polarization.  Since the modification here is a
mere unitary transformation of the Hamiltonian, or a change in the
relative phase of the Wannier functions corresponding to the different
sites within a unit cell, no new phases are found, the phase diagram
is unaltered.  If the extended SSH model is studied in the ``usual''
basis, where the phases of the Wannier functions are the same, then
there is no torus, the $d_x,d_y,d_z$ curve is an ellipse, and its
winding number around the $d_z$ axis defines the polarization (the
distinct phases have winding numbers zero or one).  When the
Hamiltonian is transformed into the basis which diagonalizes the
position, and the Brillouin zone is extended, the $d_x,d_y,d_z$ curve
becomes a knot on a torus.  Within each phase the torus on which the
``curve'' lives is of a different type, a ring or a spindle torus.
The quantum phase transition occurs when the torus is a horn torus.

These conclusions fit into the recent findings of Bena and
Montambaux~\cite{Bena09} and those of Watanabe and
Oshikawa~\cite{Watanabe18}.  The unitary transformation which relates
different representation choices can change quantities which depend on
the phase of the wave fuction.  In Ref. \cite{Watanabe18} this is
shown for the current, in our case it is shown for the polarization
(Berry-Zak phase), and for the $d_x,d_y,d_z$ curves of the system.
However, as argued in both Ref. \cite{Bena09} and \cite{Watanabe18},
phase independent quantities~\cite{Watanabe18} will not change.  The
phase independent quantity we investigated was the charge pumped in an
adiabatic process, and indeed it does not depend on the representation
used.  The path of the polarization maximum during the process,
however, does depend on it.

As for experimental tests of our work, the model suggested here can be
realized in a setting of cold atoms trapped in optical
lattices~\cite{Atala13,Nakajima16}. We anticipate that a partial
change in the polarization, of the type between maxima in
Fig. \ref{fig:phastrans} can be measured in experiments in which the
polarization is changed, as long as the process does not constitute a
full adiabatic cycle, for example, a polarization switch.  One
possible way to realize the distance dependence explicitly is the
following.  For instance a $p=1$, $q=3$ system could be realized by
constructing first a tri-partite periodic lattice, and approach our
bi-partite system as a limit by decreasing the coupling of one of the
sites in each unit cell to zero.

\section*{Acknowledgments}  BH would like to thank J. K. Asbóth and
B. Dóra for helpful discussions.  BH was supported by the National
Research, Development and Innovation Fund of Hungary within the
Quantum Technology National Excellence Program (Project Nr.
2017-1.2.1-NKP-2017-00001) and by the BME-Nanotechnology FIKP grant
(BME FIKP-NAT). \\

\end{document}